\documentclass[]{aa}  
\usepackage{graphicx}
\usepackage{txfonts}
\usepackage{subfigure}
\usepackage{natbib}
\usepackage{longtable}

\begin{document}
\newcommand{\hoA}{o-H$_2^{18}$O 1$_{10}$-1$_{01}$}
\newcommand{\hoB}{o-H$_2^{17}$O 1$_{10}$-1$_{01}$}
\newcommand{\hoC}{o-H$_2$O 1$_{10}$-1$_{01}$}
\newcommand{\hoD}{p-H$_2$O 2$_{11}$-2$_{02}$}
\newcommand{\hoE}{p-H$_2$O 2$_{02}$-1$_{11}$}
\newcommand{\hoF}{p-H$_2^{18}$O 2$_{02}$-1$_{11}$}
\newcommand{\hoG}{o-H$_2^{18}$O 3$_{12}$-3$_{03}$}
\newcommand{\hoH}{o-H$_2$O 3$_{12}$-3$_{03}$}
\newcommand{\hoI}{p-H$_2^{18}$O 1$_{11}$-0$_{00}$}
\newcommand{\hoJ}{p-H$_2^{17}$O 1$_{11}$-0$_{00}$}
\newcommand{\hoK}{p-H$_2$O 1$_{11}$-0$_{00}$}
\newcommand{\hoL}{o-H$_2$O 2$_{21}$-2$_{12}$}
\newcommand{\hoM}{o-H$_2^{17}$O 2$_{12}$-1$_{01}$}
\newcommand{\hoN}{o-H$_2$O 2$_{12}$-1$_{01}$}
\newcommand{\hoO}{p-H$_2^{18}$O 3$_{13}$-3$_{20}$}
\newcommand{\hoP}{p-H$_2$O 5$_{24}$-4$_{31}$}
\newcommand{\kms}{km~s$^{-1}$}
\newcommand{\ms}{m~s$^{-1}$}
\newcommand{\cms}{cm~s$^{-1}$}
\newcommand{\halp}{H$\alpha$}
\newcommand{\msun}{$\rm M_{\odot}$}
\newcommand{\etal}{et~al.~}
\newcommand{\vsini}{$v~sin~i$}
\newcommand{\ctemp}{$^{\circ}$C}
\newcommand{\ktemp}{$^{\circ}$K}
\newcommand{\be}{\begin{equation}}
\newcommand{\ee}{\end{equation}}
\newcommand{\bd}{\begin{displaymath}}
\newcommand{\ed}{\end{displaymath}}
\newcommand{\bi}{\begin{itemize}}
\newcommand{\ei}{\end{itemize}}
\newcommand{\bfig}{\begin{figure}}
\newcommand{\efig}{\end{figure}}
\newcommand{\bc}{\begin{center}}
\newcommand{\ec}{\end{center}}
\newcommand{\hii}{{H\scriptsize{II}}}
\newcommand{\vlsr}{V$_{\mathrm{LSR}}$}
\newcommand{\vtur}{V$_{\textrm{\tiny{tur}}}$}
\newcommand{\vexp}{V$_{\textrm{\tiny{exp}}}$}
\newcommand{\vinfall}{V$_{\textrm{\tiny{infall}}}$}
\newcommand{\coa}{$^{12}\mathrm{CO}$}
\newcommand{\cob}{$^{13}\mathrm{CO}$}
\newcommand{\coc}{$\mathrm{C}^{18}\mathrm{O}$}
\newcommand{\lsun}{L$_{\odot}$~}
\newcommand{\lfir}{L$_{\textrm{\tiny{FIR}}}$}
\newcommand{\agua}{$X_{\textrm{\tiny{H$_2$O}}}$}
\newcommand{\ratioop}{o$/$p}
\newcommand{\ratiosept}{$X_{\textrm{\tiny{$^{18}$O$/$$^{17}$O}}}$}
\newcommand{\ratiohuit}{$X_{\textrm{\tiny{$^{16}$O$/$$^{18}$O}}}$}
\newcommand{\watersept}{H$_2^{17}$O}
\newcommand{\waterhuit}{H$_2^{18}$O}
\newcommand{\water}{H$_2^{16}$O}
\newcommand{\lsol}{L$_\odot$\,}
\newcommand{\Msol}{M$_\odot$\,}

\def\etal{et al.$\;$}
 
% units of measurement

\def\kms{km\thinspace s$^{-1}$}
\def\Lsun{L$_\odot$}
\def\Msun{M$_\odot$}
\def\ms{m\thinspace s$^{-1}$}
\def\percc{cm$^{-3}$}
\title{First detection of THz water maser in NGC7538-IRS1 with SOFIA and new 22 GHz e-MERLIN maps}

\titlerunning{First detection of THz water maser in NGC7538-IRS1 with SOFIA and new 22 GHz e-MERLIN maps}

\subtitle{}

\author{F. Herpin,
	\inst{1}
	\and
	A. Baudry
	\inst{1}
	\and
	A.M.S. Richards
	\inst{2}
	\and
	M.D. Gray
	\inst{2}
	\and
	N. Schneider
	\inst{3}
	\and
	K. M. Menten
	\inst{4}
	\and
	F. Wyrowski
	\inst{4}
	\and
	S. Bontemps
	\inst{1}
	\and
	R. Simon
	\inst{3}
	\and
	H. Wiesemeyer
	\inst{4}
}

\institute{
Laboratoire d'astrophysique de Bordeaux, Univ. Bordeaux, CNRS, B18N, all\'ee Geoffroy Saint-Hilaire, 33615 Pessac, France.
	\email{fabrice.herpin@u-bordeaux.fr}
\and
JBCA, School of Physics and Astronomy, Univ. of Manchester, Oxford Road, M13 9PL, UK,
\and
I. Physik. Institut, University of Cologne, Z\"ulpicher Str. 77, 50937 Cologne, Germany,
\and
Max-Planck-Institut f\"ur Radioastronomie, Auf dem H\"ugel 69, 53121 Bonn, Germany
}

\date{\today}

\abstract
   {The formation of massive stars (M$>$10 M$_{\odot}$, L$>10^3$ L$_{\odot}$) is still not well understood. Accumulating a large amount of mass infalling within a single entity in spite of radiation pressure is possible if, among several other conditions, enough thermal energy is released. Despite numerous water line observations, over a broad range of energies, with the Herschel Space Observatory, in most of the sources observations were not able to trace the emission from the hot core around the newly forming protostellar object. }  %leave it empty if necessary
   {We want to probe the physical conditions and water abundance in the inner layers of the host protostellar object NGC7538-IRS1 using a highly excited H$_2$O line. Water maser models predict that several THz water masers should be detectable in these objects. We therefore aim to detect for the first time the o-H$_2$O $8_{2,7}- 7_{3,4}$ line in a star forming region, which model calculations predict to show maser action. }
   {We present SOFIA  observations of the o-H$_2$O $8_{2,7}-7_{3,4}$ line at 1296.41106 GHz and a $6_{16}-5_{23}$ 22 GHz e-MERLIN map of the region (first-ever 22 GHz images made after the e-MERLIN upgrade). In order to be able to constrain the nature of the emission - thermal or maser - we use near-simultaneous observations of the 22 GHz water maser performed with the Effelsberg radiotelescope and e-MERLIN. A thermal water model using the RATRAN radiative transfer code is presented based on HIFI pointed observations. Molecular water abundances are derived for the hot core.}
   {The o-H$_2$O $8_{2,7}- 7_{3,4}$ line is detected toward NGC7538-IRS1 with one feature at the source velocity (-57.7 \kms) and another one at -48.4 \kms. We propose that the emission at the source velocity is consistent with thermal excitation and is excited in the innermost part of the IRS1a massive protostellar object's closest circumstellar environment. The other emission is very likely the first detection of a water THz maser line, pumped by shocks due to IRS1b outflow, in a star-forming region. Assuming thermal excitation of the THz line, the water abundance in NGC7538-IRS1's hot core is estimated to be $5.2\times10^{-5}$ with respect to H$_2$.}
   {}
   \keywords{Stars: formation -- ISM: UCH{\sc ii} regions --masers-- molecules--individual objects: NGC7538}
%   \titlerunning{Water in G327.3--0.6}
   \maketitle
%
%________________________________________________________________

\section{Introduction}

Massive stars, despite their rarity, are important constituents of galaxies and produce most of their luminosity. The complex and still not clearly defined evolutionary sequence ranges  from massive pre-stellar cores, high-mass protostellar objects (HMPOs), hot molecular cores, and finally to the more evolved Ultra Compact HII (UCHII) region stage,  where the central object begins to ionize the surrounding gas \citep[e.g.,][]{beuther2007,koenig2017}. The classification adopted above is of course not unique. The molecular material in star-forming regions has a large range of temperatures and number densities ($10-2000$ K and $10^4-10^9$ cm$^{-3}$), with different chemical characteristics. Water  is a key molecule for determining the physical and chemical structure of star-forming regions because of large abundance variations between warm and cold regions. In cool molecular clouds, water is mostly found as ice on dust grains, but at temperatures $T>$100 K the ice evaporates, increasing the gas-phase water abundance by several orders of magnitude \citep[][]{fraser2001,aikawa2008}. At temperatures above $\sim 250$ K, all gas-phase free oxygen may be driven into H$_2$O, increasing its abundance to $\sim 3\times10^{-4}$ \citep[][]{vandishoeck2011}. 

Major progress in our understanding of interstellar water and star formation has been made with the Herschel Space Observatory\footnote{Herschel is an ESA space observatory with science instruments provided by European-led Principal Investigator consortia and with important participation from NASA.}. The guaranteed-time key program WISH \citep[Water In Star-forming regions with Herschel,][]{vandishoeck2011} probed massive-star forming regions using water observations. To collapse, the gas must be able, among several other conditions, to release enough thermal energy; a major WISH goal was to determine how much of the cooling of the warm regions ($T>$100 K) is due to H$_2$O. In particular, the dynamics of the central regions has been characterized using the water lines and the amount of cooling provided has been measured. Herschel observations of many water lines made with the high-spectral resolution Heterodyne Instrument for the Far Infrared \citep[HIFI,][]{deGraauw2010} allowed observers to perform some kind of tomography of the whole protostellar environment  and, hence, to probe the physical conditions and estimate the water abundance from a few 100 AU to a few 10000 AU from the star \citep[see][]{herpin2016}. Nevertheless, the observational evidence for water abundance jumps close to the forming stars that heat up their environs are still scarce \citep[e.g.][]{vandertak2006,chavarria2010,herpin2012,herpin2016}. While the outer abundance for the HMPOs studied in the WISH program is well constrained and estimated to be a few $10^{-8}$ in all sources, the inner abundance varies from source to source between 0.2 and 14 $\times10^{-5}$ \citep[][]{herpin2016}. Observing water lines involving high enough energy levels will allow us to probe the physical conditions and water abundance in the inner layers of the protostellar environment and thus to address this problem. One main difficulty is that up to now it has been impossible to (i) spatially resolve the region where the water jump occurs and (ii) to probe this region with the help of high excitation, optically thin lines. 

Since the end of the Herschel mission, only the Stratospheric Observatory for Infrared Astronomy \citep[SOFIA\footnote{This work is based in part on observations made with the NASA/DLR Stratospheric Observatory for Infrared Astronomy (SOFIA). SOFIA is jointly operated by the Universities Space Research Association, Inc. (USRA), under NASA contract NAS297001, and the Deutsches SOFIA Institut (DSI) under DLR contract 50 OK 0901 to the University of Stuttgart},][]{young2012} allows us to observe water in the THz frequency range. Interestingly, some of the THz water lines could even be masing. The H$_2$O excitation model of \citet{neufeld1991} as well as the works of \citet{yates1997} and, more recently, of \citet{daniel2013} and \citet{gray2016} which incorporate the most recent collisional excitation rates of water with ortho- and para-H$_2$, all make important predictions of masing transitions in the supra-THz region. Several transitions could be masing that have frequencies within the L1 and L2 channels of the German Receiver for Astronomy at Terahertz Frequencies \citep[GREAT\footnote{GREAT is a development by the MPI f\"ur Radioastronomie and the KOSMA$/$Universit\"at zu K\"oln, in cooperation with the MPI f\"ur Sonnensystemforschung and the DLR Institut f\"ur Planetenforschung},][]{heyminck2012} on SOFIA \citep[][]{heyminck2012}. In particular, three predicted o-H$_2$O masing transitions and one p-H$_2$O masing transition fall in the 1.25-1.50 THz range. Only one o-H$_2$O transition, $8_{2,7} - 7_{3,4}$ at 1296.41 GHz, could be observed at the flying altitude of the SOFIA observatory (transmission $\simeq$ 62\%), the other ones being absorbed by the remaining atmosphere. Emission from this transition involves an upper energy level at 1274.2 K, and can probe the gas deep enough into the inner layers of the hot core. We note that \citet{neufeld2017} have just reported the first detection of the $8_{2,7} - 7_{3,4}$ masing line toward an evolved star. 

We have selected NGC7538-IRS1, the best studied massive object from our WISH program \citep[][]{herpin2016,vandertak2013}, which has already been observed in the 22 GHz maser line and exhibits strong thermal water lines. NGC7538-IRS1 is a relatively nearby \citep[2.65 kpc,][]{mosca2009} UCHII object in the complex massive star-forming region NGC7538 surrounded by a molecular hot core. The 22 GHz H$_2$O masers associated with it exhibit a complex spatial distribution and the strongest and main concentration of the maser features \citep[see][]{kameya1990,surcis2011} is found in the direction of IRS1. Untill recently \citep[see][]{beuther2013} the central source was thought to be an O6 star (30 \Msol, $L\simeq 8\times 10^4$ \lsol) forming one single high-mass Young Stellar Object (YSO). New interferometric observations of the methanol masers \citep[][]{mosca2014} have demonstrated that NGC7538-IRS1 actually consists of three individual high-mass YSOs named IRS1a, IRS1b, and IRS1c within 1600 AU. Note that IRS1a, b, and c are associated with the methanol maser clusters labeled by \citet{minier2000} B+C, A, and E, respectively. The most massive YSO is IRS1a, with $\sim$25 \Msol and a quasi-Keplerian disk of $\sim$1 \Msol which dominates the bolometric luminosity of the region. Another massive ($\leq 16$ \Msol) and thick disk orbits around the less massive (a few \Msol) IRS1b object. The third source, IRS1c, is likely to be a massive YSO too. 

Several bipolar outflows or jets emanating from IRS1 region in multiple directions have been identified and characterized. A North-South free-free ionized jet with an opening angle $\leq 30^{\circ}$ has been observed by \citet{sandell2009} as well as a strong accretion flow toward IRS1 (\.M$\sim2\times 10^{-3}$ \Msol$/$yr). Several studies have characterized the NW-SE \citep[$PA=-50^{\circ}$,][]{qiu2011} CO outflow now thought to originate from IRS1a \citep[][]{mosca2014}. In addition, a NE-SW outflow with $PA=40^{\circ}$ has been observed by \citet{beuther2013}. An outflow driven by IRS1b and collimated by its rotating disk has also been observed \citep[][]{mosca2014}. The presence of jets$/$outflows and strong accretion flows makes this shocked region an ideal place for exciting masers. 

In this paper we report an observational study of the NGC7538-IRS1 region. The o-H$_2$O $8_{2,7}- 7_{3,4}$ emission was observed with the GREAT instrument onboard SOFIA. In addition, the 22 GHz water maser emission was observed using the Effelsberg telescope\footnote{The Effelsberg 100-m telescope is a facility of the MPIfR (Max-Planck-Institut f\"ur Radioastronomie) in Bonn.} and e-MERLIN interferometer\footnote{e-MERLIN is a national facility operated by The University of Manchester on behalf of the Science and Technology Facilities Council (STFC)}. In section 2 we describe the observations, and in section 3, we present the observational data. In section 4 we discuss the nature of the detected THz water emission. In section 5 we discuss the region kinematics and physical conditions, and in section 6 we summarize the results of our study.

\section{Observations and data reduction}

\subsection{SOFIA observations}

Observations in singlepoint chopping mode of the $8_{2,7}-7_{3,4}$ line of ortho-\water~were carried out toward NGC7538-IRS1 ($\alpha_{{\rm J} 2000} = 23^{\rm h}13^{\rm m}45.3^{^s}, \delta_{{\rm J} 2000} = +61^{\circ}$28'10.0") as part of SOFIA Cycle 3 project on 2015 December 9, using GREAT. At the systemic velocity of NGC7538-IRS1, the system temperature was typically 2000 K (SSB) and signal-band zenith opacity 0.08. The on-source integration time was 20 min.  The chop throw was 100\arcsec to either side of the on-source position (chop-nod method). One channel (L1) of GREAT was tuned to the 1296.41106 GHz water line frequency (lower sideband LSB), the other channel, the LFA (Low Frequency Array) 7 pixel array, was tuned to the [CII] 158 $\mu$m line. The [CII] line was detected but a discussion of these results will be presented elsewhere. We employed the digital 4GFFT spectrometer \citep[][]{klein2012} that analyzed a bandwidth of 1.5 GHz with a spectral resolution of 0.283 MHz (0.056 \kms). Data have later been smoothed to 1.129 \kms. The rms of the spectra is 90 mK at this latter spectral resolution. 

Data have been processed with the latest version of the GREAT calibrator and converted from the $T_{A}^*$ scale ($\eta_{f}=0.97$) into main beam temperature units, $T_{MB}$, applying the main beam coupling efficiency $\eta_{mb} =0.69$ for the L1 channel. The conversion factor Jy/K to be applied on the T$_{A}^*$ data is 971 Jy$/$K. The half-power beam width is 20.6\arcsec at this frequency. All spectra have been calibrated for the transmission in the signal band and the continuum level correction for double-sideband reception has been applied. Further analysis was done within the CLASS\footnote{http://www.iram.fr/IRAMFR/GILDAS/} package. The continuum level (SSB) is $T_{MB}$ = 2.0 K (at 1296.4 GHz).

\subsection{Effelsberg observations}

In order to constrain the maser models, and because of the variability of the maser emission, nearly contemporaneous observations of the $6_{16}-5_{23}$ transition of ortho-H$_2$O (rest frequency 22.23508 GHz) were carried out on 2015 December 11 (integration time on source of 15 min) with the MPIfR 100-m telescope at Effelsberg, Germany. The half power beamwidth was $\simeq39$\arcsec~and the pointing accuracy was, on average, 4\arcsec. We used the 1.3 cm double beam secondary focus receiver (K$/$Jy conversion factor of 1.6) with the XFFT backend in high resolution mode (32768 channels  with a spectral resolution of $\sim$0.04 \kms). Data have been smoothed to 0.1 \kms~ and the rms noise at this resolution is 37mJy. System temperature was 190 K. All data have been calibrated in Jy; the calibration parameters were derived by continuum observations of suitable flux density calibrators. The main beam efficiency is 0.64. The calibration uncertainty is about 10-15\%. 

\begin{figure}
\centering
\includegraphics[width=8.6cm, angle=0]{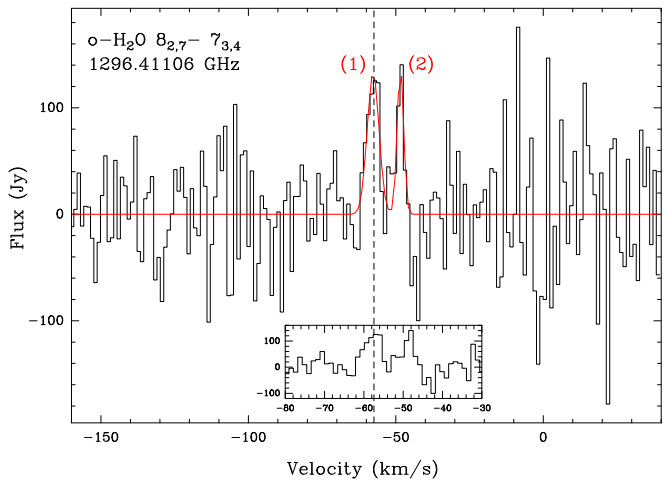}
\caption{Continuum-subtracted SOFIA spectra of the o-H$_2$O $8_{2,7}- 7_{3,4}$ line at 1296.41106 GHz toward NGC7538-IRS1. The vertical dotted line indicates \vlsr~ at -57.4 \kms. The spectral resolution is 1.129 \kms. The Gaussian fit is shown in red. The small insert shows the same spectra on a zoomed velocity scale.}
\label{SOFIA}
\end{figure}

\subsection{e-MERLIN observations}

Interferometric observations of the same source of the ortho-H$_2$O 6$_{16} - 5_{23}$ line emission were performed with e-MERLIN (commissioning observations) on 2016 April 20.  These were the first-ever 22 GHz images made after the e-MERLIN upgrade. Four telescopes were used, the longest baseline being 217 km, with the 25-m Mark 2, Darnhall and Pickmere telescopes and the 32-m Cambridge telescope.
 At the assumed distance of 2.7 kpc, the synthesized beam of 20 mas corresponds to a spatial resolution element of approximately 54 AU, but the positions of individual bright maser components can be fitted with a relative accuracy $\sim1$ AU. 

 The observations were made in full Stokes parameters (although only the parallel bands have been processed, giving total intensity images). One spectral window (spw) of 4 MHz was centred on the  22.23508 GHz maser line (adjusted for the source velocity immediately prior to observations), divided into 512 channels giving a spectral resolution of 0.105 km s$^{-1}$, total useful span $\sim$50 km s$^{-1}$. Two similar spw were placed either side (but turned out not to contain any emission).  A single 125 MHz spectral window (also with 512 channels) was placed overlapping these for calibration.  The pointing position of $\alpha_{{\rm J} 2000} = 23^{\rm h}14^{\rm m}01.749^{^s}, \delta_{{\rm J} 2000} = +61^{\circ}$27'19.80" was used for NGC 7538, which was approximately observed 10 hr on target, in 6-min scans interleaved with the phase reference source J2302+6405. The quasar 3C84 was used to provide the bandpass correction and flux scale, with a flux density in 2016 April of 37 Jy (measurements kindly provided by L\"{a}hteenm\"{a}ki, Metsahovi, private communication). The LSR (Local Standard of Rest) correction was applied to the line data before self-calibration and imaging.  

\begin{figure}
\centering
\includegraphics[width=8.6cm, angle=0]{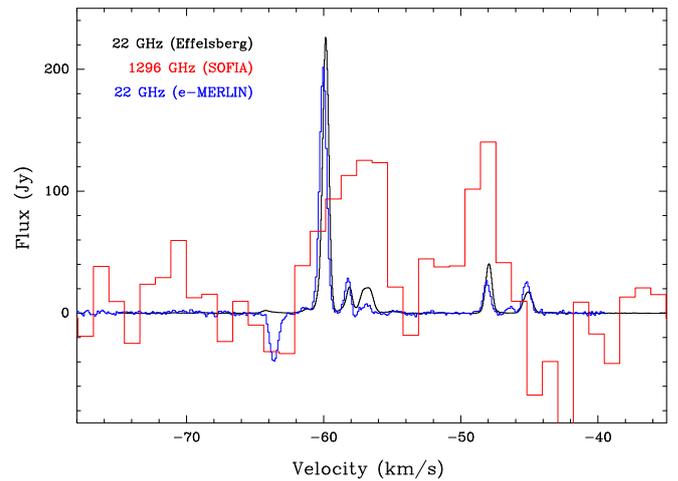}
\caption{Effelsberg (in black, spectral resolution $\delta$v of 0.1 \kms) and e-Merlin (in blue, $\delta$v=0.1 \kms, the apparent absorption near -63.5 \kms~is attributed to e-MERLIN residual sidelobes, see Sect. \ref{map_discussion}) spectra of the $6_{16}-5_{23}$ transition of ortho-H$_2$O at 22 GHz toward NGC7538-IRS1 plotted over the SOFIA ($\delta$v=1.1 \kms) spectrum of the o-H$_2$O $8_{2,7}- 7_{3,4}$ line at 1296.41106 GHz (in red).}
\label{comp_spectres}
\end{figure}

The calibration was performed in CASA, and final imaging and component fitting in AIPS.  The  position of the brightest maser was established as a reference position before self-calibration. Test image cubes were made at coarse resolution to look for masers within $\sim$1.5 arcmin radius of the pointing centre (within the half-power point of the primary beam). Four fields were imaged at full resolution, covering all the emission identified, see Table~\ref{tab:imparms}.

\begin{table*}
\caption{e-MERLIN 22 GHz imaging parameters: centre of square field, length of side, fractional error in flux scale, primary beam (PB) correction} 
\label{tab:imparms}
\begin{tabular}{llrrr}
\hline
Field & RA, Dec. (J2000)            &Size & Flux error &PB correction\\
      & hh:mm:ss dd:mm:ss           &$\arcsec$& \%&divisor\\
\hline
New   &23:13:43.69442 +61:27:49.3597& 6   & 6 & 0.95\\
IRS11 &23:13:44.98183 +61:26:49.6800&  6  & 10&0.88\\
IRS1-3&23:13:45.14200 +61:28:11.4100& 12  & 5&1.0\\
IRS9  &23:14:01.75100 +61:27:19.8000& 6   & 15&0.8\\
\hline
\end{tabular}
\end{table*}

The e-MERLIN beam at 22 GHz is not yet fully characterised. We made images without primary beam correction, and then fitted 2D Gaussian components to each patch of emission above 4 $\sigma_{\mathrm{rms}}$ (rejecting components isolated in fewer than 3 channels or any obvious sidelobes of brighter components).  We estimated the primary beam corrections using a scaled VLA beam, and divided the measured fluxes by the factors given in Table~\ref{tab:imparms},  with uncertainties increasing with distance from the pointing centre. The most distant fields were also slightly affected ($<10\%$) by time smearing.

The noise $\sigma_{\mathrm{rms}}$ is $\sim20$ mJy in quiet channels for fields close to the pointing centre (IRS 1-3), rising to double that in the most distant fields. The sparse visibility plane coverage means that the brighter channels are severely dynamic-range limited, with $\sigma_{\mathrm{rms}}$ typically $10\%$ of peaks, or even more in more remote fields. The astrometric position accuracy is a few tens milli-arcsec (mas); the relative accuracy depends on the signal to noise ratio and is $<1$ mas for all but the faintest or most remote components.

\subsection{HIFI observations}

NGC7538-IRS1 was part of the WISH GT-KP sample. Fifteen water lines (see~Table \ref{table_transitions}) were observed with the HIFI spectrometer on-board Herschel Space Observatory in the pointed (or mapping)  mode at frequencies between 547 and 1670 GHz in 2010 and 2011 (list of observation identification numbers, {\em obsids}, are given in Table \ref{table_transitions}) toward NGC7538-IRS1 (same coordinates as for SOFIA observations). For the pointed observations, the Double Beam Switch observing mode with a throw of $3'$ has been used. The off positions have been inspected and do not show any emission. The frequencies, energy of the upper levels, system temperatures, integration times, and {\it rms} noise levels at a given spectral resolution for each of the lines are provided in Table \ref{table_transitions}. 

\begin{figure*}
\centering
\includegraphics[width=17.cm, angle=0]{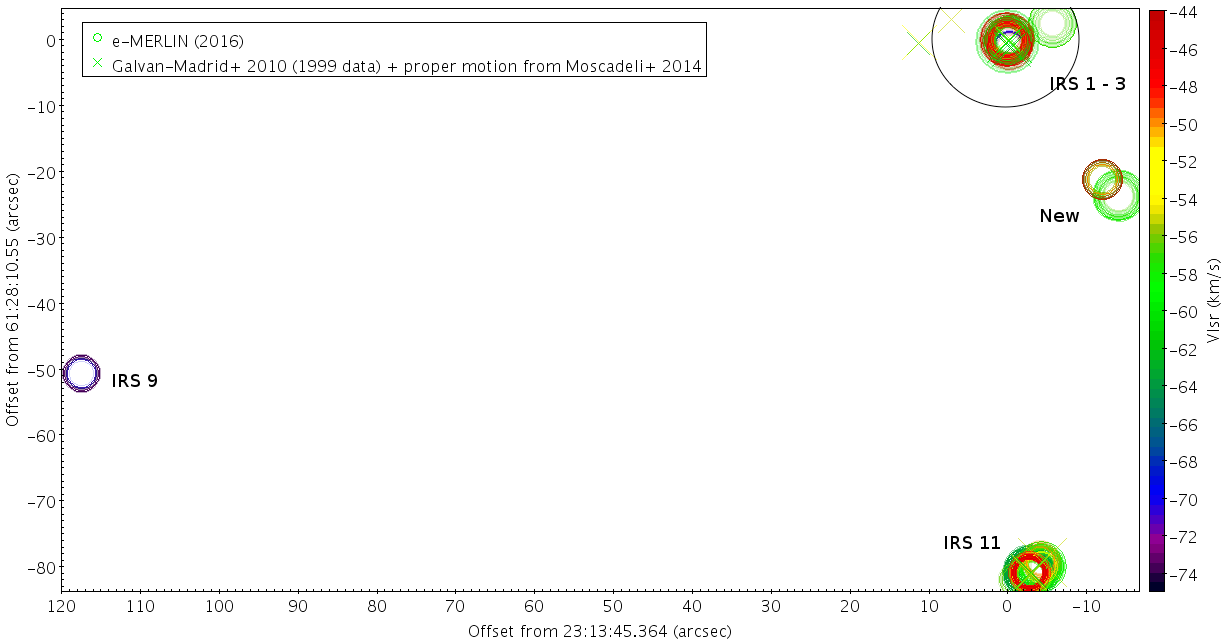}
\caption{Entire e-MERLIN map of 22 GHz maser emission from NGC7538, including IRS1-3, IRS9 and IRS11, of the 22 GHz water maser line. Colored circles show the relative position of individual maser features, with color denoting the maser $V_{LSR}$, according to the color-velocity conversion code reported on the right side of the panel. The water maser positions reported by \citet{Galvan2010} and corrected for proper motion \citep[][]{mosca2014} are shown as crossesn(same color-velocity conversion code) for comparison. The beam of the SOFIA observation is overlaid in black for comparison.}
\label{Map_total}
\end{figure*}
\begin{figure*}
\centering
\includegraphics[width=15.cm, angle=0]{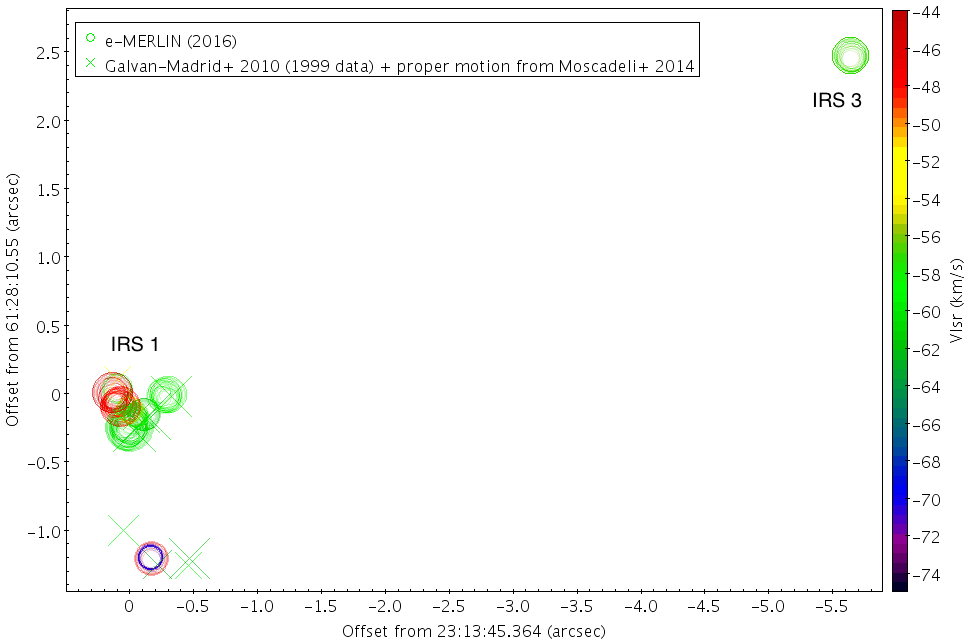}
\caption{As in Fig. \ref{Map_total} but zoomed on IRS1-3.}
\label{Map_total_zoom_IRS1}
\end{figure*}

Data were taken simultaneously in H and V polarizations using both the acousto-optical Wide-Band Spectrometer (WBS) with 1.1 MHz resolution and the digital auto-correlator or High-Resolution Spectrometer (HRS). Calibration of the raw data into the $T_A$ scale was performed by the in-orbit system \citep[][]{roelfsema2012}. Conversion to $T_{mb}$ was done using the beam efficiency\footnote{http://herschel.esac.esa.int/twiki/pub/Sandbox/TestHifiInfoPage/} given in Table~\ref{table_transitions} and a forward efficiency of 0.96. The flux scale accuracy is estimated to be between 10\% for bands 1 and 2, 15\% for band 3, 4, and 20 \% in bands 6 and 7$^1$. Data calibration was performed in the Herschel Interactive Processing Environment \citep[HIPE,][]{ott2010} version 14. Further analysis was done within  CLASS. These lines are not expected to be polarized. Thus, after inspection, data from the two polarizations were averaged together. Because HIFI is operating in double-sideband, the measured continuum level has been divided by a factor of 2 in the Figures and the Tables (this is justified because the sideband gain ratio is close to 1). Note that the \hoP line is suffering from strong baseline and ripples problems.

\section{Analysis}
\subsection{The THz water emission}

The ortho-\water~$8_{2,7}-7_{3,4}$ line emission is detected toward NGC7538-IRS1 with SOFIA. As shown in Fig. \ref{SOFIA}, the line profile exhibits two features at -57.7$\pm0.6$ and -48.4$\pm0.5$ \kms~ where 22 GHz emission is also observed (see Fig. \ref{comp_spectres}). From Gaussian fitting the S$/$N ratio of the integrated lines at -57.7 and -48.4 \kms~is 5.5 and 3.2, respectively. Hereafter the -57.7 and -48.4 components will be named features (1) and (2), respectively. At the frequency of the observed emissions there is no contamination by other molecular species according to {\em Splatalogue} catalog\footnote{http://www.cv.nrao.edu/php/splat/}. The peak intensities and linewidths are 133.1 Jy and 4.7 ($\pm1.3$) \kms~for feature (1), and 139.4 Jy and 2.6 ($\pm1.3$) \kms~for feature (2).

The velocity and linewidth of feature (1) are comparable to what has been observed from the CS thermal lines \citep[$v_{LSR}=-57.4$ \kms, $\delta v=5.5\pm1$ \kms,][]{vandertak2000a} or OH with SOFIA \citep[]{csengeri2012}. On the contrary, feature (2) is centered at a velocity different from the generally adopted source velocity. In addition the line is narrower by a factor of 2. The nature of these water components is discussed in Sect. \ref{nature}.

\begin{figure*}
\centering
\includegraphics[width=17.cm, bb = 50 180 564 669, angle=0]{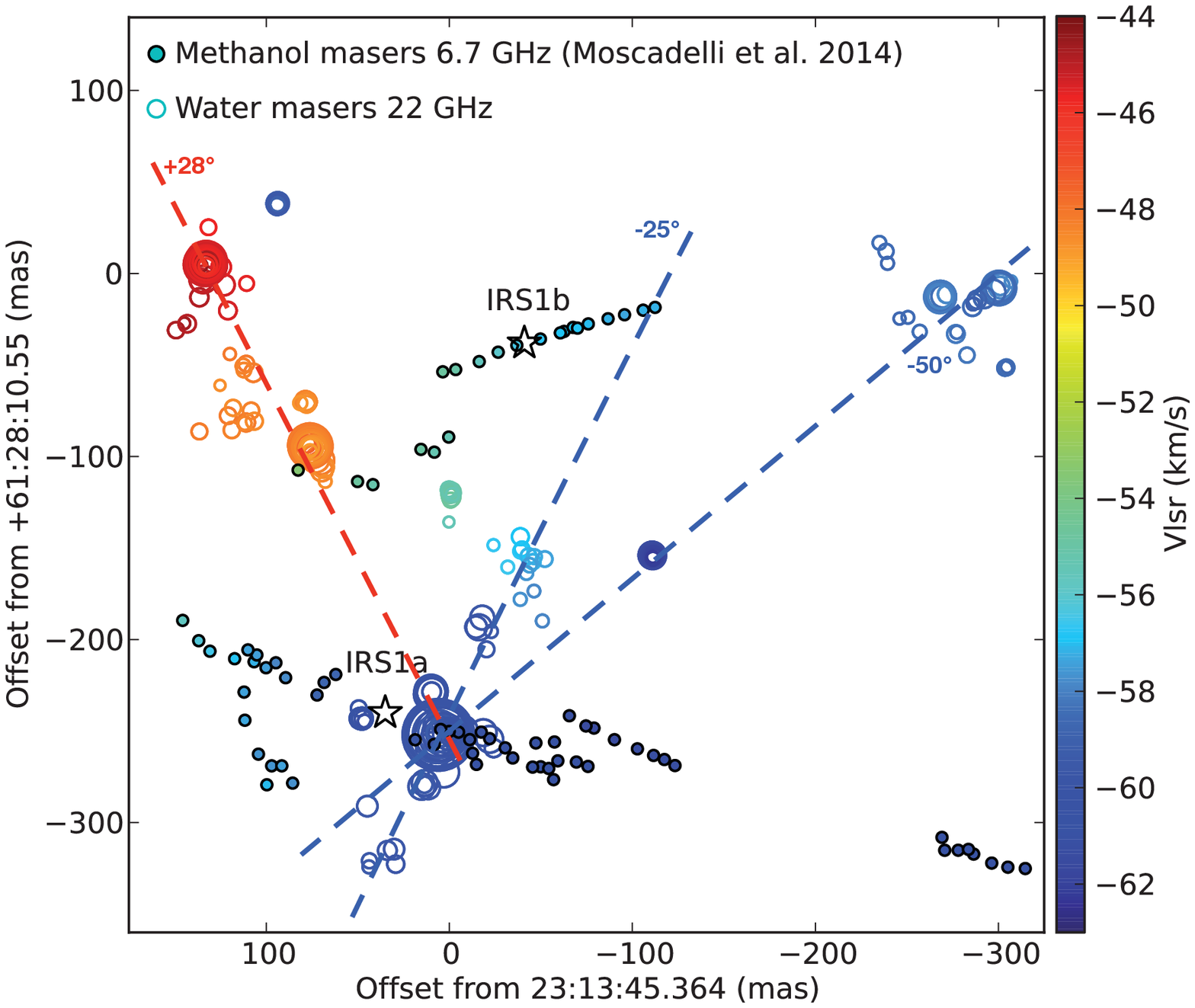}
\caption{Zoom towards IRS1 North of the e-MERLIN map shown in Fig. \ref{Map_total}. Colored empty  and filled circles show the relative position of individual 22 GHz water masers (this paper) and 6.7 GHz methanol maser features \citep[][]{mosca2014}, respecively. $V_{LSR}$ is according to the color-velocity conversion code shown on the right side of the panel.}
\label{Map_total_zoom_IRS1_Mosca}
\end{figure*}

\subsection{The 22 GHz single-dish observation}

Several velocity components of the 22 GHz water maser line are seen in the Effelsberg spectra (see Fig. \ref{comp_spectres}). All lines are narrow with a width of the order of 1 \kms~or less. Two main groups of lines can be distinguished: one made of three strong lines close to the source velocity ($\sim-(56-60)$ \kms)) and another one with two lines centrered at -45.1 and -48 \kms. These lines are highly variable with time (on a several months timescale) as shown by \citet{felli2007} from their long-term monitoring observations. Nevertheless, most of the time the $\sim-60$ \kms~feature is the strongest maser line. 

According to the existing 22 GHz maser maps and our e-MERLIN map (see next Section) all of these maser components are associated to IRS1-3.  The non-detections of more distant masers are due to the fact that the Effelsberg telescope's primary beam is less that 1$/$3 that of e-MERLIN.

\subsection{Map of the 22 GHz maser emission}
\label{map_discussion}

The entire e-MERLIN map, shown in Fig. \ref{Map_total}, encompasses the highly luminous infrared sources IRS1-3, IRS9, and IRS11 which were first identified by \citet{werner1979}. We have identified 286 individual features which can be gathered into 109 groups based on their peak velocity and coordinates. All these groups are reported in Table \ref{table_water_pos} where their position, peak flux density $S_{\nu}$, peak velocity, and velocity range are given. We have compared our maser positions with the work of \citet{Galvan2010}, taking into account the proper motions from \citet{mosca2014}, $\mu_{RA}$ = -2.45 mas$/$yr and $\mu_{DEC}$ = -2.45 mas$/$yr. Both maps are consistent. Our Figs. \ref{Map_total} and \ref{Map_total_zoom_IRS1} around IRS1-3 and IRS11 show maser positions very close to those labeled M and S in Table 2 of \citet{Galvan2010} and shown here as green crosses. Moreover, the total size of the image around IRS1-3 (12\arcsec, a size comparable to the SOFIA beam at 1.3 THz) has been used to synthesize the e-MERLIN 22 GHz spectrum which we compare on Fig. \ref{comp_spectres} with the line profile obtained at Effelsberg in order to verify the flux density scale (e-MERLIN data are commissioning data), and to check that the interferometer is detecting all the flux. Both line profiles are very similar except from the apparent negative feature in the -63.5 to -64 \kms~velocity range due to residual sidelobes arising from strong IRS11 emission in this velocity range. The maser variability over this short time period (4 months) is thus not significant and all the flux is detected. 

\begin{table*}
%\begin{minipage}[t]{500pt}
  \caption{Observed line emission parameters for the detected lines with HIFI toward NGC7538-IRS1. $T_{mb}$ is the peak temperature with continuum. $T_{cont}$ is the real continuum temperature (single sideband). $\varv$ is the Gaussian component peak velocity. $\Delta\varv$ is the velocity full width at half-maximum (FWHM) of the narrow, medium, and broad components. The opacity $\tau$ is from absorption lines.}
\begin{center}
\label{table_param}      % is used to refer this table in the text
\begin{tabular}{lccccccccc} \hline \hline
{\bf Line}  & $T_{mb}$ & $T_{cont}$ & $\varv_{nar}$ & $\Delta\varv_{nar}$ & $\varv_{med}$ & $\Delta\varv_{med}$ & $\varv_{br}$ & $\Delta\varv_{br}$  & $\tau$ \\ 
 & [K] & [K] & [\kms]  & [\kms] & [\kms] & [\kms] &  [\kms] &[\kms]  & \\ 
\hline                        
   \hoA     &  0.63 & 0.5  &  & &  -58.4$\pm$0.4& 5.2$\pm$1.1& & & \\ 
   \hoF     &  1.90 & 1.7 & $-59.5\pm0.2$ & $3.9\pm0.5$&  & & & & \\ 
   \hoG   & 2.27 & 2.0  & & & -58.2$\pm$0.2& 5.1$\pm$0.3& &  &  \\
   \hoI   & 2.22 & 2.0  & & &  -57.7/-64.3$^a$$\pm$0.2& 5.1/8.8$\pm$0.2 &  & & 0.03\\
   \hoJ    & 1.9 & 2.0  & -56.7$\pm0.3^a$& 3.2$\pm0.6$&  & &  &  & 0.04\\
%   \hoM   & 4.95 & 5.6  & -43.2$\pm$0.2$^a$ & 3.1$\pm$0.4 & -40.1$\pm0.9^a$& 10$\pm2$ & -54$\pm3^a$& 21$\pm2$ & 0.12$\pm$0.05 \\
\hline
   \hoC   & 5.21 & 0.5  & -57.1$^a$/-54.7$^a$$\pm0.2$ & 2.0/3.9$\pm$0.2 & & & -57.2$\pm$0.2 & 13.0$\pm$0.3 & 0.9\\
   \hoD      & 6.86 & 0.8 & -57.5$\pm0.1$& 4.3$\pm0.1$ & &  & -57.6$\pm$0.1& 15.2$\pm$0.3&\\
   \hoP$^b$      & 1.75 & 1.6  & & &  &  & -57.6$\pm$0.7& 12.2$\pm$1.6  & \\
   \hoE      & 7.92 & 1.6  & -57.4$\pm$0.1 & 4.4$\pm$0.1& &  & -57.7$\pm$0.1 & 12.3$\pm$0.3 & \\
  \hoH   & 6.32 & 2.0  & -57.3$\pm0.1$ & 3.4$\pm0.1$ & -57.8$\pm$0.1& 8.5$\pm$0.2 &  &  &\\
   \hoK$^c$    & 0.1 & 2.0  & -56.1$\pm0.4^a$ & 4.6$\pm0.1$ & -51.2$\pm$0.4$^a$ & 9.3$\pm$0.6 & -55.8$\pm$0.2 & 24.0$\pm$0.4 & 2.9\\
   \hoL     & 5.9 & 4.0$^d$  & & &  -60.6$^a$/-56.3$\pm$0.3 & 5.1/8.7$\pm$0.5 &  & & 0.4\\ 
   \hoN    & 0.0 & 4.0$^d$ & &  & -58.2$^a$/-52.1$\pm$0.3& 5.7/8.1$\pm$0.5& & & $>$7\\
\hline
\end{tabular}
\end{center}
\tablefoot{$^a$ in absorption, $^b$ Line suffering from strong baseline and ripples problems, $^c$ WBS data, $^d$ the continuum level is not reliable}
%\end{minipage}
\end{table*}

We detect 68 maser groups for IRS11, some of them very bright  with $S_{\nu}$ well above 100 Jy, spanning a velocity range from -44.75 to -65.08 \kms. The maximum flux is observed at -60 \kms~(964 Jy) in IRS1. Only three main groups of masers are observed toward IRS9 and they are blue-shifted ($\sim-(69-79)$ \kms) compared to IRS1 source velocity. The strongest maser group peaks at 39 Jy in IRS9.  

Approximately 25 arcseconds south-west from IRS1 (offsets $\sim$-(12-14)$\arcsec$ and -(21-24)$\arcsec$), a new source has been found exhibiting powerful maser emission (up to $\sim$200 Jy for group 5) and velocities from -45 to -63 \kms~(see Fig. \ref{Map_total} and Table \ref{table_water_pos}).  The feature "E" observed by \citet{kameya1990} with $V_{LSR}=-63.3$ \kms~at  $\alpha_{{\rm J} 2000} = 23^{\rm h}13^{\rm m}42.51^{^s}, \delta_{{\rm J} 2000} = +61^{\circ}$27'45.1", i.e. at offset $\sim-20\arcsec$ and $-25\arcsec$, is much further from this new spot.

A first zoom of the IRS1-3 region is shown in Fig. \ref{Map_total_zoom_IRS1}. The IRS3 source lies to the NW with offsets respective to IRS1 of -5.5$\arcsec$ and 2.5$\arcsec$ in RA-DEC, respectively. Two main groups are detected for IRS3, they are moderately bright (14-88 Jy) and span a small velocity range (from $\sim$ -56 to -58 \kms). At this scale, IRS1 is made of what \citet{surcis2011} called the {\em North source} (the main one) and the {\em South} source, $1.2\arcsec$ south. In addition to the maser group at $\sim-70$ \kms~for IRS1, already observed by these authors, we detect maser spots at redder velocities $\sim-46$ \kms. 

The close-up view of \water~maser features around NGC7538-IRS1 North (see Fig. \ref{Map_total_zoom_IRS1_Mosca}) reveals 26 maser groups which can be split into two distinct velocity ranges: a first large group of masers within $\pm3$ \kms~from the main source LSR velocity, represented in blue on Fig. \ref{Map_total_zoom_IRS1_Mosca}, and a second group of red-shifted  masers whose LSR velocity is between -45 and -48 \kms. The "blue" group is made of numbers of features concentrated very close to IRS1a \citep[as defined by][]{mosca2014} plus several spots spread North-West. The "red" group masers lying North of IRS1a are less aligned along a NE-SW direction. The maser spatial distribution is discussed in more detail in Sect. \ref{spatial}.

\section{Nature of the THz water emission}
\label{nature}

\subsection{Spatial origin}
\label{spatial}

Considering the SOFIA beam size ($\simeq$ 21\arcsec) and the approximate 12\arcsec~square size of the image made by e-MERLIN (primary beam, see Table. \ref{tab:imparms}) towards IRS1-3  (see Fig. \ref{Map_total} and \ref{Map_total_zoom_IRS1}), the detected THz water emission can only originate from IRS1-3. Then, comparing the velocity range of features (1) and (2) with the velocity of maser spots detected by e-MERLIN (Figs. \ref{comp_spectres} and \ref{Map_total_zoom_IRS1} and Table \ref{table_water_pos}), we can infer that:
\begin{itemize}
   \item water in IRS1 North or IRS3 can give rise to feature (1),
  \item only IRS1 North exhibits 22 GHz maser emission at velocities ($\sim -48.4$ \kms) similar to feature (2) THz emission. 
\end{itemize}

At the scale of Fig. \ref{Map_total_zoom_IRS1_Mosca}, we can see that sizes of the respective emitting regions for maser features around -48 and -58 \kms are comparable, but not spatially coincident. Maser spots at velocities similar to  THz feature (2) emission are located 0.2\arcsec~North-East of the high-mass YSO IRS1a and even more than 0.1\arcsec~East of the massive protostar IRS1b. We also conclude that a similar beam dilution of both maser features is obtained with the Effelsberg 22 GHz beam ($\simeq$ 39 \arcsec). 

\subsection{Thermal or maser?}

\subsubsection{Thermal modeling}
\label{sec:model}

\begin{figure*}
   \begin{minipage}[c]{0.46\linewidth}
     \includegraphics[width=8.cm]{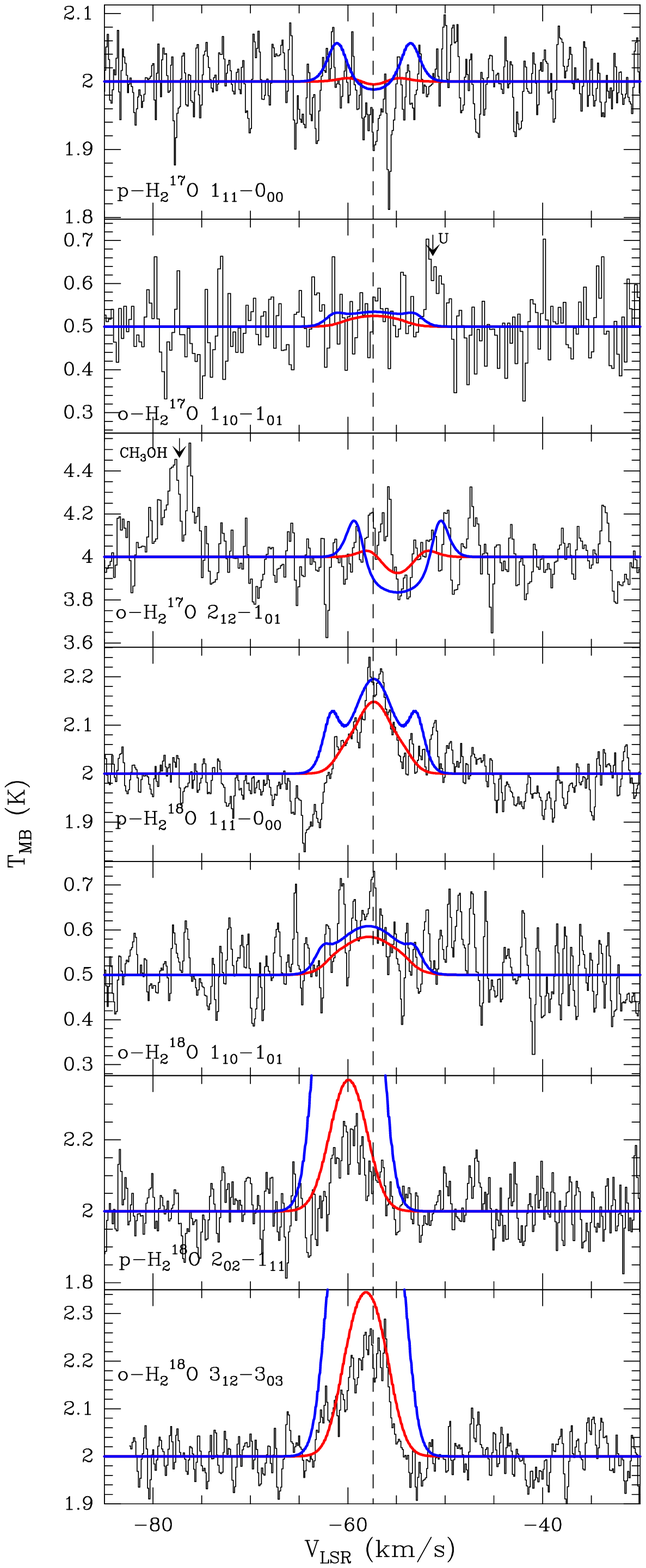}
   \end{minipage} \hfill
   \begin{minipage}[c]{1.96\linewidth}
      \includegraphics[width=8cm]{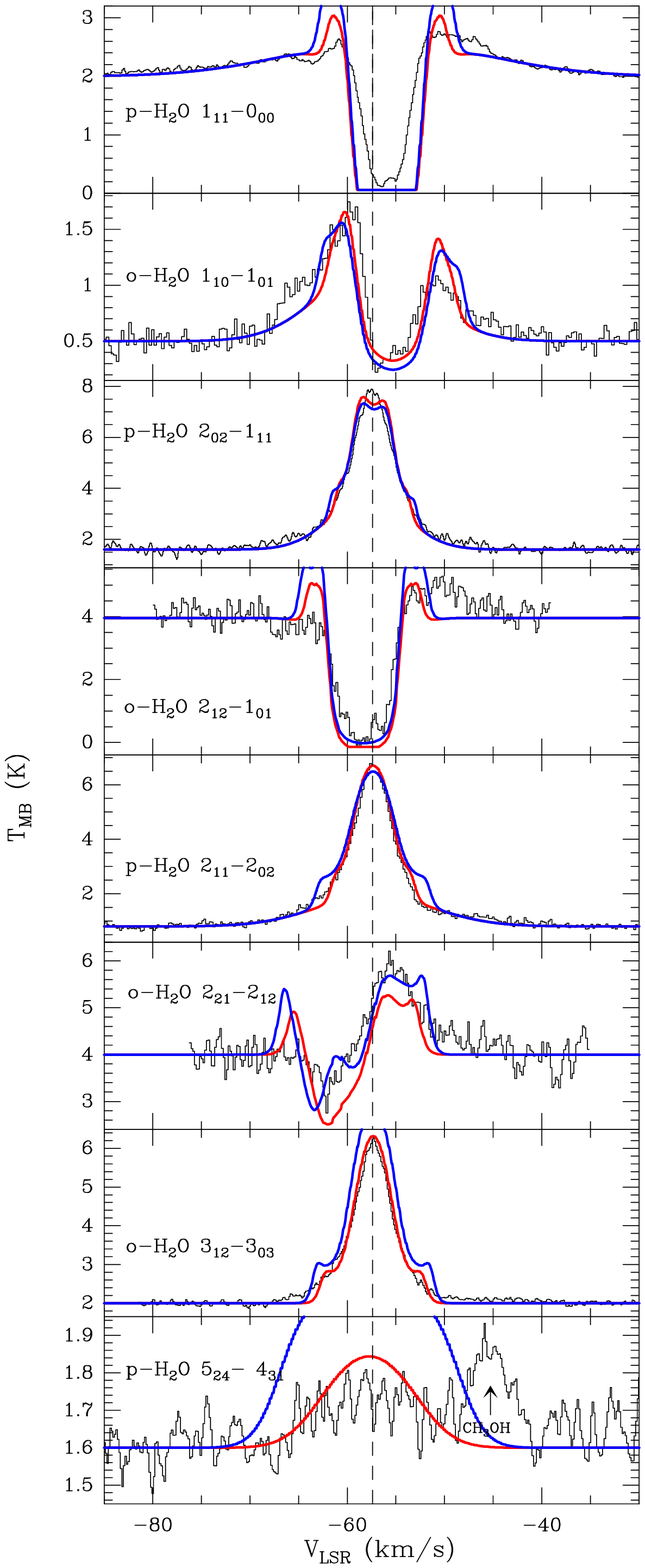}
   \end{minipage}
\caption{HIFI spectra of \watersept$/$\waterhuit~({\em Left}) and \water~({\em Right}) lines (in black) respectively, with continuum for NGC7538-IRS1 pointed position. The best-fit model is shown as red line over the spectra ($\chi_{in}$(H$_2$O)$=8\times10^{-6}$ and $\chi_{out}$(H$_2$O)$=4\times10^{-8}$). The model adopting the {\em SOFIA} water inner abundance ($5.2\times10^{-5}$) is shown as blue line over the spectra. Vertical dotted lines indicate the \vlsr~(-57.4 \kms~from the line modeling). The spectra have been smoothed to 0.2 \kms, and the continuum divided by a factor of two.}
\label{FigHIFI}%
\end{figure*}

All water lines observed toward NGC7538-IRS1 with HIFI are shown in Fig. \ref{FigHIFI}. A detailed analysis of three low-J \water~line profiles has been made by \citet{vandertak2013} in several HMPOs including NGC7538-IRS1. Several lines from the rare isotopologue \waterhuit~and the para-ground-state line of \watersept~have been detected in addition to the \water~lines towards NGC7538-IRS1 (see Fig. \ref{FigHIFI}, left).

A Gaussian fit to these line profiles (see Table \ref{table_param}) indicates, depending on the line, three velocity components: a narrow (3.2-4.6 \kms), a medium (5.1-9.3 \kms), and a broad ($> 12.2$ \kms) velocity component. The physical origin of these components (attributed to a dense core, an envelope, and an outflow) has been discussed in several papers \citep[e.g.][]{herpin2016}. All of these components in NGC7538-IRS1 are globally centered at a velocity similar to that of the THz feature (1), i.e. $\sim -57$ \kms. No emission is detected at -48 \kms, which is thus a first indication of the non-thermal origin of this emission. 

\begin{figure}
\centering
\includegraphics[width=9.cm, angle=0]{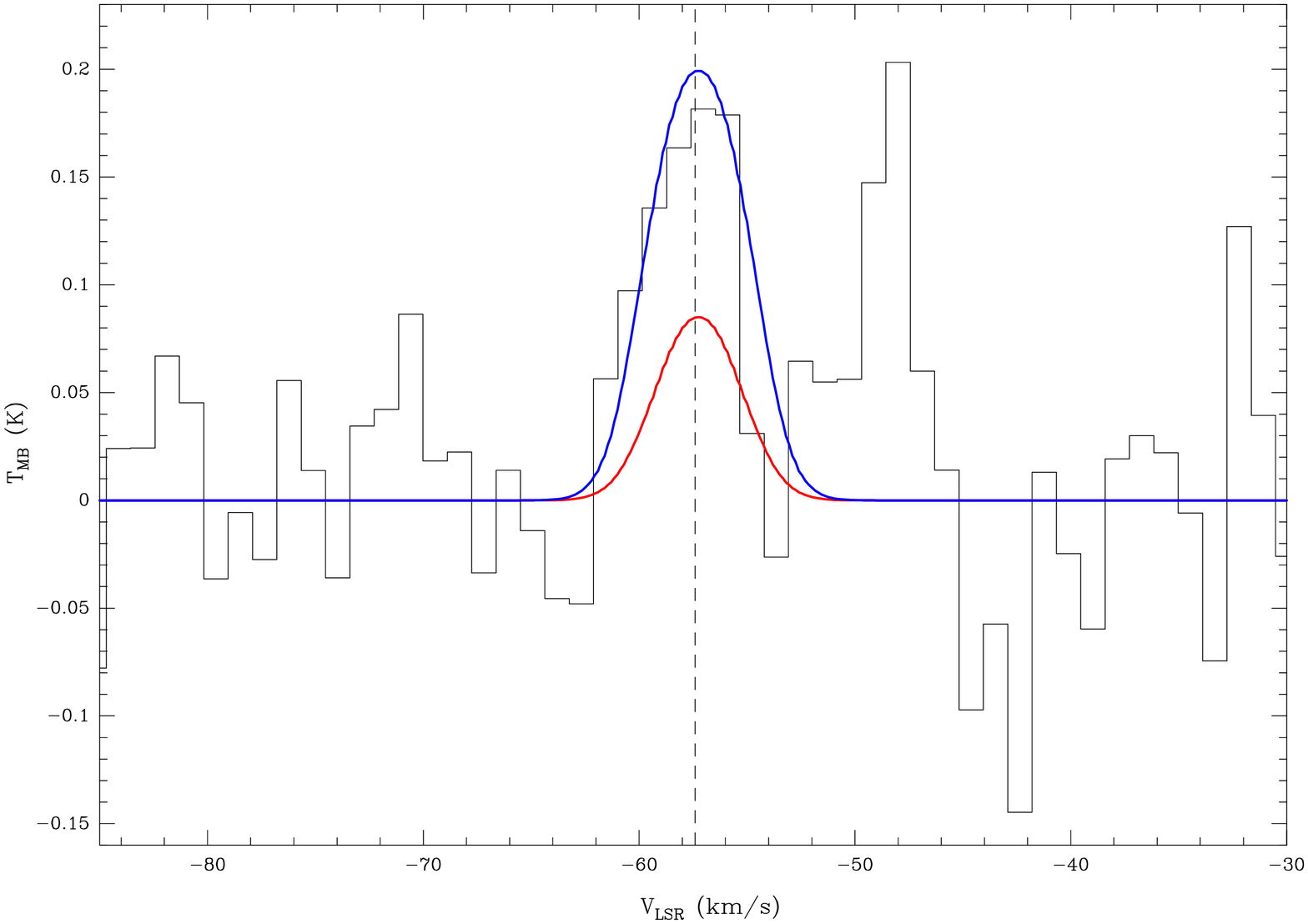}
\caption{SOFIA spectra of the o-H$_2$O $8_{2,7}- 7_{3,4}$ line showing for feature (1) the best-fit thermal model in red from our HIFI data (see Fig. \ref{FigHIFI}) and a model adopting a water inner abundance of $5.2\times10^{-5}$ (blue profile).}
\label{HIFImodel_SOFIA}
\end{figure}

Following the method of \citet{herpin2012,herpin2016}, we have modeled all water line profiles in a single spherically symmetrical model using the 1D-radiative transfer code RATRAN \citep[][]{hogerheijde2000}. The envelope temperature (20-1130 K) and  number density ($1.1\times 10^5-3.8\times 10^8$ cm$^{-3}$) structure for the hot core is from \citet{vandertak2013}. This analysis assumes a single-source within the HIFI beam, hence encompassing the IRS1-3 substructure. The outflow parameters, intensity and width, come from our Gaussian fit (see Table \ref{table_param}) for the broad component. The envelope contribution is parametrized with the water abundance (outer $\chi_{out}$ for $T<100$ K, inner $\chi_{in}$ for $T>100$ K, assuming a jump in the abundance in the inner envelope at 100 K due to the evaporation of ice mantles), the turbulent velocity ($v_{turb}$), and the infall velocity ($v_{inf}$). We adopt the following standard abundance ratios for all the lines: 4.5 for \waterhuit$/$\watersept~\citep[][]{thomas2008}, 614 for \water$/$\waterhuit~\citep[based on][]{wilson1994}, and 3 for ortho$/$para H$_2$O.  

Considering that the width of the velocity components is not the same for all lines (see Table \ref{table_param}), a model with equal turbulent velocity for all lines does not fit the data well. The best result (see Fig. \ref{FigHIFI}) is obtained with a turbulent velocity of 1.5-2.5 \kms~ (5.5 \kms~ is even needed for the doubtful \hoP~line) for \water, depending on the modeled line, and 2.5 \kms~for the rare isotopologues lines. No infall is needed at the scale probed by the HIFI lines. All modeled lines are centered at roughly $-57.4\pm 0.5$ \kms. 

Modeling of the entire set of observed lines constrains the water abundance, but only a few lines are optically thin enough to probe the inner part of the envelope. We hence derive \water~abundances relative to H$_2$ of $8\times 10^{-6}$ in the inner part and of $2\times 10^{-8}$ in the outer part. 

We have then applied the results of the above thermal model to feature (1) of the THz o-H$_2$O $8_{2,7}- 7_{3,4}$ line (we do not see any thermal line corresponding to feature 2), assuming the same abundance and $v_{turb}=2.5$ \kms. Figure \ref{HIFImodel_SOFIA} shows that we do not reproduce the observed line, the intensity being only half what is observed. Actually, we can perfectly reproduce feature (1) with an increased water inner abundance of $5.2 \times 10^{-5}$ (see blue line in Fig. \ref{HIFImodel_SOFIA}). We have also successfully applied this new model to the \waterhuit~$3_{13}-2_{20}$ line ($E_{up}\simeq$200 K) reproducing the line intensity and width observed by \citet{vandertak2006}.  

Adopting now the water inner abundance deduced from our modeling of the o-H$_2$O $8_{2,7}- 7_{3,4}$ line, we have modeled again the HIFI water thermal lines (in blue on Fig. \ref{FigHIFI}). The result is less satisfactory, specially for the \hoP (but this observation suffers from line ripples and the baseline determination is also affected by methanol line blending) and rare isotopologue lines. This might reflect the limitations of our symmetrical 1-D model. It is also possible that the $8_{2,7}- 7_{3,4}$ line detected by SOFIA emanates from the inner part of the hot core whilst the water lines observed with HIFI are from somewhat cooler regions further out. Thus, a temperature gradient could explain our results without requiring a higher water abundance throughout the region. On the other hand, feature (1) of the 1296 GHz water line and the \waterhuit~$3_{13}-2_{20}$ line can be well  matched by a thermal emission profile if the water abundance is increased. Nevertheless, non-thermal effects cannot be excluded: our RATRAN model could miss a non-thermal contribution at 1296 GHz which, in that case, would be on the order of 50\%.

 \begin{figure}
\centering
\includegraphics[width=8.6cm, angle=0]{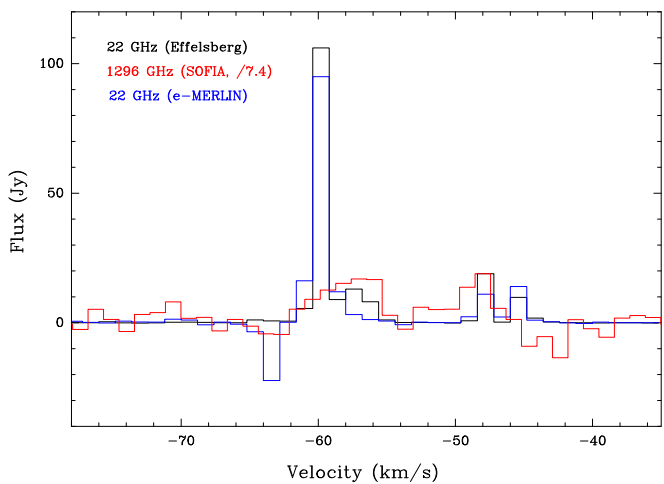}
\caption{As in Fig. \ref{comp_spectres} with all spectra smoothed to a spectral resolution of 1.1 \kms. The SOFIA fluxes have been divided by a factor 7.4. }
\label{comp_spectres_smooth}
\end{figure}

\subsubsection{Maser modeling}

We have made quasi-contemporaneous observations of the potentially masing o-H$_2$O $8_{2,7}-7_{3,4}$ transition at 1296 GHz and of the 22 GHz maser line to provide line intensity ratios enabling us to estimate the physical conditions leading to these maser emissions. Line intensity ratios are less dependent on the cloud geometry, and in the saturated regime such ratios tend to be independent of the exact ratio of the beaming angles which should then be close to one. 

We estimate the 1296/22 opacity ratio from the recently published models of \citet{gray2016}. These models incorporate 411 levels and 7597 radiative transitions, the most recent collision rates, and line overlap effects. Even if these models have been done primarily for evolved stars, the broad physical parameter space (including $T_{dust}$ rising from 50 K to 350 K) that has been explored can be used for the conditions considered here. There is a significant overlap between conditions supporting the 22 GHz and 1296 GHz masers, e.g. both could come from gas above 1000 K with o-H$_2$O number densities between $10^4$ and $2\times10^5$ cm$^{-3}$ \citep[i.e. $n_{H_2} \sim  3\times 10^9-6\times 10^{10}$ cm$^{-3}$, taking $3\times 10^{-5}$ as standard conversion factor from $n$(H$_2$O) to $n$(H$_2$),][]{gray2016}. However, strong 1296 GHz maser emission is obtained for a relatively narrow range of physical conditions, typically $T_K$ around 1000 to 3000 K and $n_{H_2O}$ around $10^4-10^5$ cm$^{-3}$, while 22 GHz masers are excited within 500 to 3000 K and for a broader range of H$_2$O densities up to several $10^5$ cm$^{-3}$. Compared to the 22 GHz inversion, the 1296 GHz inversion is biased toward  higher values of $T_K$ and it is lost with increasing dust temperatures, namely a low value of $T_{dust}$, around 50 K, is preferred. Actually, the dust temperature is unlikely to be much above 50 K in shocked gas. Assuming $T_{dust}$ = 50 K, \citet{gray2016} predict $\tau(1296)/\tau(22)\sim1.9$. This ratio does not result from specific conditions representing the observed source. It is, instead, the ratio of the maximum 1296 GHz depth to the maximum found at 22 GHz from models covering a large parameter space. Saturation is not accounted for in this ratio either. Nevertheless, the 1.9 ratio suggests that the flux density and brightness temperature at 1296 GHz can be several times that at 22 GHz.
	To compare features (1) and (2) to our model predictions we derive the line peak ratios from Fig. \ref{comp_spectres_smooth} where all spectra have been smoothed to the same spectral resolution. We derive S(1296)$/$S(22) = 1.2 and 7.4 for features (1) and (2), respectively, but cannot directly estimate the brightness temperature of the 1296 GHz emission since its spatial extent is unknown.  At -48 \kms, feature (2) is definitely much brighter in the 1296 GHz line than at 22 GHz. This is in agreement with maser conditions suggested from the 1.9 opacity ratio above. We further note that feature (2) is nearly 2.5 times narrower than feature (1) suggesting again that the former is not thermal. Assuming that the 1296 and 22 GHz emissions have a similar spatial extent we derive from the observed S(1296)$/$S(22) = 7.4 at -48 km/s, $T_b(1296) = T_b(22) \times 2.2\times10^{-3}$ and, because the condition $T_b(22)$ of order of or greater than $10^6$ K is easily met at 22 GHz \citep[e.g.,][]{elitzur1992}, we get $T_b(1296)$ greater than 2200 K, i.e. suprathermal emission. We stress that deriving an exact value for $T_b(1296)$ is impossible since we have no spatial information. But using the e-MERLIN resolution of 20 mas as an upper limit to the beamed size of maser spots gives $T_b$ in the range $>10^6$ to $>10^9$ K, with a mean value of $>3\times10^7$ K. In fact, the beamed area of maser spots is likely to be at least two orders of magnitude smaller, giving an average $T_b$ at least $10^9$ K. This lower limit to $T_b$ strongly suggests, but does not prove, the maser nature of this emission. 
%densities between $\sim 10^4$ and $2\times 10^5$ cm$^{-3}$. Actually strong 1296 GHz maser will form under a narrower range of physical conditions than the 22 GHz one. The 1296 GHz inversion is not only biased towards higher kinetic temperatures compared to 22 GHz but is also lost with increasing dust temperatures, i.e. a low value of $T_{dust}$ around 50 K is preferred. These conclusions might be slightly changed by applying a stronger IR pumping field. Hence, the prediction from \citet{gray2016} gives a typical opacity ratio $\tau(1296)/\tau(22)\sim1.9$, suggesting a brightness ratio of 6.7.
%
%From our observation (see Fig. \ref{comp_spectres_smooth}) we derive a line peak ratio (1296$/$22) of 1.2($\pm$0.2) and 7.4($\pm$1.5) for features (1) and (2), respectively. While the ratio for feature (1) is very different from what is expected for maser emission, feature (2) water intensity is compatible with predictions from \citet{gray2016}. Of course, we assume here that the 22 GHz maser distribution applies to the 1296 GHz maser components, if present. The small difference between the predicted and measured maser ratios could be due to one maser being amplified over a greater column density and$/$or a larger angular distribution (within the beam). 

Our conclusion is thus that feature (2) water emission is likely a maser. The absence of typical narrow substructure in the line profile could just be due to the low S$/$N (and spectral resolution) hiding it. Observing with some instrument that has much higher spatial resolution at 1296 GHz would be the only way to definitely prove it is a maser.

\section{Discussion}

\subsection{Water content}

Depending on the excitation of the o-H$_2$O $8_{2,7}- 7_{3,4}$ line, purely thermal or not, we have shown that the water inner abundance might differ by more than one order of magnitude in NGC7538-IRS1's hot core. From the integrated fluxes measured for the lines (considered in this study) at least partly in emission, we have derived the water luminosities, and then have estimated, assuming isotropic radiation, that the minimum total HIFI water luminosity is 0.6 L$_{\odot}$ (equal to the sum of all individual observed luminosities). Even if the true water emission from the inner parts might be much greater, the cool envelope absorbs much of the emission. This confirms the low contribution of water cooling to the total far-IR gas cooling compared to the cooling from other species \citep[][]{karska2014,herpin2016}. Moreover, from the modeling, and assuming that feature (1) water emission is purely thermal, we estimate the total water mass in the envelope to be $10^{-3}$ \Msol~and that 93\% of this mass resides in the inner parts, to be compared with $2\times10^{-4}$ \Msol~and 69\% in the case of a non-thermal contribution. 

\subsection{Kinematics and Geometry of the region}

The previous section suggests that the THz water feature (1) is consistent with thermal excitation (even if a non-thermal contribution is possible) while feature (2) could be masing. Different physical conditions, i.e. a different spatial origin, could explain these different behaviours. According to \citet{gray2016}, this could be for instance explained by a higher dust temperature in the region where the -57 \kms~water is excited compared to the -48 \kms~region (the o-H$_2$O $8_{2,7}- 7_{3,4}$ inversion is lost with increasing dust temperature). 

When plotting over our 22 GHz e-MERLIN map the 6.7 GHz methanol maser spots and the location of high-mass YSO IRS1a and b from \citet{mosca2014}, the -48 \kms~22 GHz maser spots appear located close to the IRS1b source while the -57 \kms~ones are associated to the IRS1a source. Interestingly, \citet{mosca2014}, based on NH$_3$ maps, explain that the gas surrounding IRS1b (a less evolved source) has a lower temperature than that observed toward IRS1a which is a more massive and evolved YSO, with $T$ lower than 250 K. Hence, higher temperature in IRS1a might be less suitable for water maser emission at 1296 GHz. But more likely IRS1a could collisionally quench the maser if the number density is too high. We have estimated the critical density (at $T=50$ K) corresponding to the 1296 GHz transition to be $5\times 10^7$ cm$^{-3}$. The gas in the core of IRS1a is thus so dense (number densities up to a few $10^8$ cm$^{-3}$ can be reached in the inner part, see Sect \ref{sec:model}) that it begins to quench the maser action.

The maser spot distribution as derived from our e-MERLIN map provides new information when compared to previous published works. Figure \ref{Map_total_zoom_IRS1_Mosca} shows the spatial location of the 22 GHz maser features and the 6.7 GHz methanol masers of \citet{mosca2014}. According to these authors, the two individual high-mass YSOs, IRS1a and b, lie at the center of a line of methanol masers tracing disks with position angles of $PA=+71^{\circ}$ and $-73^{\circ}$, respectively. The strongest "blue" (i.e. $v\leq -56$ \kms) maser spots we detect can be associated to IRS1a, others being distributed roughly along a line with $PA=-25^{\circ}$ and another with $PA=-50^{\circ}$. Some similar linear distribution with $PA=-52^{\circ}$ was seen by \citet{surcis2011} who suggested that these water masers were almost aligned with the CO NW-SE outflow, elongated 0.3 pc from IRS1a with $PA=-40^{\circ}$ \citep[][]{kameya1990,gaume1995,qiu2011}. They proposed the water masers are pumped by a shock caused by the interaction of the outflow with the infalling gas \citep[observed at scales $\geq$1000 AU,][]{beuther2013}. Actually, the e-MERLIN map shows water masers following this NW-SE axis of the CO outflow but at slower velocities compared to what \citet{qiu2011} observed in CO (-78;-64 \kms). These H$_2$O maser spots might either trace the cavity of the outflow, i.e. a cone with an opening angle of $\sim25^{\circ}$ and $PA\simeq-40^{\circ}$, or trace two different outflows originating from IRS1a, the outflow at $PA=-25^{\circ}$ being almost perpendicular  to the disk. 

Maser spots whose velocity is close to the THz feature (2) (i.e. -44;-50 \kms) are located NW from IRS1a and W from IRS1b. They are distributed also along a line with $PA\simeq+28^{\circ}$, i.e. NE-SW, which can be associated with the outflow observed by \citet{beuther2013} in HCO$^+$(4-3) with $PA\simeq40^{\circ}$. Moreover, the NH$_3$ \citep[][]{mosca2014}, OCS, CH$_3$CN, and $^{13}$CO \citep[][]{zhu2013} observations show a velocity gradient in the same direction ($PA\sim30-40^{\circ}$) with line emission at velocities similar to our feature (2) emission in this NE region. \citet{mosca2014} explain this "red-shifted" NH$_3$ features toward the NE by an outflow driven by IRS1b and collimated by its rotating disk. Hence, as proposed above on the basis of a too high temperature towards IRS1a, we propose that the THz feature (2) is a maser, not associated with IRS1a, and pumped by shocks driven by the IRS1b outflow. 

\section{Conclusions}
\label{sec:Conclusions}

SOFIA observations toward NGC7538-IRS1 of the o-H$_2$O $8_{2,7}- 7_{3,4}$ line emission are presented. Two separate velocity features are detected, one is associated with the source velocity (-57.7 \kms), and another one lies at -48.4 \kms. Combining these observations with near-simultaneous observations of the $6_{1,6}- 5_{2,3}$ masing transition of ortho-H$_2$O at 22 GHz with the Effelsberg telescope and with the e-MERLIN interferometer, we discuss the nature of these THz emission features. 

A thermal water model based on HIFI observations can reproduce the $8_{2,7}- 7_{3,4}$  line component at the source velocity if the water inner abundance is increased by more than an order of magnitude to $5.2\times10^{-5}$, compared to what is estimated by the "HIFI alone" model. In addition, the observed brightness ratio (1296$/$22) for both features is compared to the maser predictions and lead us to conclude that while the THz emission feature at the systemic velocity is mostly thermal, the -48.4 \kms~feature is  likely masing.

We argue that the two line components do not arise from the same location, i.e. different physical conditions could explain these different natures. The thermal emission is excited in the innermost part of the IRS1a protostellar massive object and is an excellent probe of the water reservoir in the inner part of the hot core. We suggest that the maser emission is associated with shocks driven by the IRS1b outflow. 

\begin{acknowledgements}
We would like to thank Anne L\"{a}hteenm\"{a}ki (Mets\"{a}hovi Radio Observatory, Aalto University, Finland) for the flux measurements of 3C84  and Alex Kraus (MPIfR-Bonn, Germany) for having performed and reduced the Effelsberg observations.  We thank the SOFIA operations and the GREAT instrument teams, whose support has been essential for the GREAT accomplishments, and the DSI telescope engineering team. Based [in part] on observations made with the NASA$/$DLR Stratospheric Observatory for Infrared Astronomy. Sofia Science Mission operations are conducted jointly by the Universitis Space Research Association, Inc., under NASA contract NAS297001, and the Deutsches SOFIA Institut, under DLR contract 50 OK 0901. {\it Herschel} is an ESA space observatory with science instruments provided
by European-led Principal Investigator consortia and with important
participation from NASA. 
HIFI has been designed and built by a consortium of
 institutes and university departments from across Europe, Canada and the
 United States under the leadership of SRON Netherlands Institute for Space
 Research, Groningen, The Netherlands and with major contributions from
 Germany, France and the US. Consortium members are: Canada: CSA,
 U.Waterloo; France: CESR, LAB, LERMA, IRAM; Germany: KOSMA,
 MPIfR, MPS; Ireland, NUI Maynooth; Italy: ASI, IFSI-INAF, Osservatorio
 Astrofisico di Arcetri- INAF; Netherlands: SRON, TUD; Poland: CAMK, CBK;
 Spain: Observatorio Astron{\'o}mico Nacional (IGN), Centro de
 Astrobiolog{\'i}a
 (CSIC-INTA). Sweden: Chalmers University of Technology - MC2, RSS $\&$
 GARD; Onsala Space Observatory; Swedish National Space Board, Stockholm
 University - Stockholm Observatory; Switzerland: ETH Zurich, FHNW; USA:
 Caltech, JPL, NHSC.). 
\end{acknowledgements}

\bibliographystyle{aa}
\bibliography{biblio}

\begin{appendix}
%\section{Intrinsic sideband ratio IF dependence}

\section{Herschel$/$HIFI observed water line transitions.}

\begin{table*}
%\begin{Large}
\caption{Herschel/HIFI observed water line transitions. The rms is the noise in $\delta \nu=1.1$MHz.}             % title of Table
\label{table_transitions}      % is used to refer this table in the text
\centering                          % used for centering table
\begin{tabular}{lccccccccc}        % centered columns (4 columns)
\hline\hline                 % inserts double horizontal lines
Water species & Frequency &  Wavelength & $E_u$    & HIFI & Beam  &  $\eta_{\textrm{mb}}$  &  $T_{\textrm{sys}}$   &  rms & obsid \\   
                      &    [GHz]     &       [$\mu$m] &      [K]   &band &  [\arcsec]  &  & [K]  & [mK] & \\
\hline                        
   \hoA$^a$    & 547.6764     &  547.4 &  60.5    & 1a & 37.8  & 0.62 & 80 & 58 & 1342202036 \\
   \hoB    & 552.0209    &  543.1  & 61.0     &  1a & 37.8 &  0.62 & 70  & 42 & 1342198332-3\\     
   \hoF     & 994.6751  &  301.4  &  100.6   & 4a & 21.1   &  0.63 & 290  & 69  & 1342197964-5\\
   \hoG    & 1095.6274  &  273.8  &  248.7   &  4b & 19.9  &  0.63 & 380   & 48  & 1342200760-1\\
   \hoI    & 1101.6982  &  272.1  &   52.9   &  4b & 19.9 &  0.63 &  390   & 36  & 1342191663-4, 1342197976\\
   \hoJ    & 1107.1669  &  272.1  &   52.9   &   4b & 19.9 &  0.63 &  380   & 37 & 1342200760-1\\
   \hoM    & 1662.4644  &  180.3  &  113.6   & 6b  & 12.7 &  0.58 & 1410   & 232  & 1342200758\\
\hline
   \hoC$^a$    & 556.9361    &  538.3  &   61.0   & 1a & 37.1  & 0.62 & 80   & 65 & 1342202036 \\
   \hoD    & 752.0332    &  398.6  &  136.9   & 2b & 28.0   &  0.64 &  90   & 67 & 1342201546-7\\
      \hoP$^b$    & 970.3150    &   309.0 &  598.8   & 4a & 21.8   &  0.63 &  620   & 59 & 1342227536\\
  \hoE    & 987.9268 &  303.5  &   100.8  & 4a & 21.3  & 0.63 &  340   & 69 & 1342197964-5\\
   \hoH    & 1097.3651  &  273.2  &  249.4   & 4b  & 19.9  &  0.63 & 380   & 49  & 1342200760-1\\
   \hoK    & 1113.3430  &  269.0  &  53.4   &4b & 19.7   &  0.63 & 395   & 36 & 1342191663-4, 1342197976\\
   \hoL    & 1661.0076  &  180.5  &  194.1   &6b & 12.7  &  0.58 & 1410   & 232 &1342200758\\
   \hoN    & 1669.9048  &  179.5  &   114.4   & 6b  & 12.6  &  0.58 & 1410   & 232 & 1342200758\\
\hline                                  
\end{tabular}
\tablefoot{$^a$ This line was mapped in OTF mode. $^b$ Line suffering from strong baseline and ripples problems.}
%\vspace{30 mm}
%\end{Large}
\end{table*}

\section{Water maser positions}
\label{sec:water_positions}

\begin{table*}
%\begin{Large}
\caption{Water Maser positions and velocities at 22 GHz (e-MERLIN observations of April 2016).}             % title of Table
\label{table_water_pos}      % is used to refer this table in the text
\centering                          % used for centering table
\begin{tabular}{lcccccc}        % centered columns (4 columns)
\hline\hline                 % inserts double horizontal lines
IRS	& Group & offset X$^a$  & offset Y$^a$  &	Peak $S_{\nu}$	 & Peak $V_{LSR}$	& Velocity Range \\   
                      &       &       [$\arcsec$] &     [$\arcsec$]  & [Jy] &  [\kms]  &   [\kms]  \\
\hline                        
1 & 1 & 0.132 &-0.010 &17.31 &-45.06 & -44.64 to -45.70 \\
1 & 2 & 0.133 &0.005 &143.23 &-45.27   &-44.43 to -45.91 \\
1 & 3 & 0.070 &-0.106 &13.12 &-48.33 &    -48.22 to -48.65 \\
1 & 4 & 0.0762 &-0.095 &145.27 &-48.12 &   -47.49 to -48.96 \\  
1 & 5 & 0.114 &-0.081 &4.21 &-48.12 &  -48.12 to -48.54 \\
1 & 6 & 0.114 &-0.061 &4.61  & -48.22 & -48.12 to -48.64 \\
1 & 7 & 0.123 &-0.060 &1.01 &-48.12 &-48.12 to -48.33 \\
1 & 8 & 0.079 &-0.070 &2.91 &-48.33 &  -48.33 to -48.75 \\
1 & 9 & -0.001 &-0.122 &9.16 &-54.96 &-54.44 to -55.49 \\
1 & 10 & -0.042 &-0.160 &23.23 &-57.91 &-56.54 to -57.91 \\
1 & 11 & -0.288 &-0.010 &29.24 &-58.23 &      -57.81 to -58.55 \\
1 & 12 & -0.287 &-0.012 &62.02  &-58.34 &-57.81 to -59.07 \\
1 & 13 & -0.278 &-0.037 &2.92 &-58.12 &-58.12 to -58.34 \\
1 & 14 & -0.304 &-0.051 &1.44 &-58.34 & -58.34 to -58.65 \\
1 & 15 & -0.251 &-0.027 &0.86 &-58.65 & -58.44 to -58.65 \\
1 & 16 & -0.238 &0.012 &1.88 &-58.44   & -58.44 to -58.65 \\
1 & 17 & 0.094 &0.038 &8.59 &-59.28 &   -58.86 to -59.49 \\
1 & 18 & 0.005 &-0.250 &112.31 &-60.13 &     -59.92 to -60.65 \\
1 & 19 & 0.006 &-0.234 &29.61 &-60.23 &     -60.13 to -60.44 \\
1 & 20 & -0.012 &-0.258 &26.90 &-60.55 &     -60.13 to -60.55 \\
1 & 21 & 0.004 &-0.250 &964.41 & -60.01 & -59.18 to -61.07  \\
1 & 22 & 0.020 &-0.2820 &10.34 &       -59.81  &-59.39 to -59.81 \\         
1 & 23 & -0.018 &-0.195 &12.34  &       -59.70  &-59.39 to -59.81  \\
1 & 24 & 0.036 &-0.319 &6.97  &       -59.70  &-59.39 to -59.81  \\
1 & 25 & 0.048 &-0.243 &6.46 &-60.76 &-60.55 to -61.07 \\
1 & 26 & -0.111 &-0.154 &27.73 &     -61.39 &-60.86 to -62.23 \\ 
1S & 1 & -0.163 &-1.201 &14.32  &   -70.66 &-69.20 to -70.98  \\
1S & 2 & -0.170 &-1.211 &27.09 &-46.43 &    -45.80 to -47.07 \\
3 & 1 & -5.647 &2.456 &14.40 &-56.86 &-56.33 to -57.60 \\
3 & 2 & -5.642 &2.474 &88.55 &-56.76 &-56.12 to -57.91 \\
9 & 1 & 117.433 &-50.752 &2.06 &-69.18 &      -68.97 to -69.19 \\
9 & 2 & 117.432 &-50.751 &2.76 &-69.39 &      -69.39 to -69.61 \\
9 & 3 & 117.436 &-50.742 &39.13& -73.61 &-72.45 to -74.98 \\
11 & 1 & -2.747 & -80.919 & 7.56 & -44.96   & -44.75 to -45.70 \\
11 & 2 & -2.736 & -80.885 &32.20 & -45.48 & -45.27 to -45.60   \\
11 & 3 & -2.743 &-80.926 & 29.41 & -47.06 & -45.91 to -47.80  \\
11 & 4 & -2.733 &-80.888 & 12.24  &-46.75 &  -46.33 to -47.60 \\
11 & 5 & -2.748 & -80.970  & 2.38 & -47.28 &-46.85 to -47.49  \\
11 & 6 & -2.738 &-80.984  &  6.84   & -47.06 & -46.64 to -47.59  \\
11 & 7 & -2.730 &-80.846 &   2.83 & -46.85 &-46.64 to -47.17 \\
11 & 8 & -2.715 &-80.960 & 1.82 & -46.85 &-46.65 to -47.06  \\
11 & 9 & -2.684 &-80.908 & 2.05 & -47.38 & -47.17 to -47.38  \\
11 & 10 & -4.127 &-79.095 &   9.53 & -50.01 & -49.59 to -50.33 \\ 
11 & 11 & -4.112 &-79.057 & 5.23 & -49.59 &-49.59 to -50.22  \\
11 & 12 & -4.116 &-79.160 &   3.88 & -50.12 & -49.91 to -50.22 \\ 
11 & 13 & -4.123 &-79.126 & 1.69 & -50.44 & -50.22 to -50.44  \\
11 & 14 & -4.300 &-80.046 &   3.94 & -53.49 & -53.28 to -53.81 \\ 
11 & 15 & -4.293 &-80.012 &   1.64 &   -53.81 & -53.38 to -53.81  \\
11 & 16 & -2.928 &-81.067  &21.02 &   -54.12 & -53.81 to -54.65  \\
11 & 17  & -2.918 &-81.120 &   11.80   & -54.44 -&53.91 to -54.65  \\
11 & 18  & -2.931 &-80.992 &   5.31  & -54.23 &-53.91 to -54.44  \\
11 & 19 & -2.919 &-81.039 &   3.24 & -54.23 &-53.91 to -54.44  \\
11 & 20 & -2.874 &-81.056 &   2.21 &  -54.33 &-54.33 to -54.54  \\
11 & 21 & -2.793 &-80.885 &   4.52  & -55.60 &-55.07 to -55.70  \\
11 & 22 & -2.784 &-80.861 &   2.92 &   -55.28 &-55.07 to -55.49  \\ 
11 & 23 & -2.914  &-81.168 &1.77 &-55.81 & -55.81 to -56.02 \\
11 & 24 & -1.121 &-81.967 &10.40 &-56.02  &-55.91 to -57.97 \\
11 & 25 & -1.112 &-81.940 &1.90 &-56.23    &-56.23 to -56.65  \\ 
\hline                                  
\end{tabular}
\tablefoot{$^a$ All offsets are relative to the position RA=23h13m45.364s, DEC= +61$^{\circ}$28'10.550"). The uncertainty is less than 0.003$\arcsec$.}
\vspace{30 mm}
%\end{Large}
\end{table*}
\setcounter{table}{0}

\begin{table*}
%\begin{Large}
\caption{continued}             % title of Table
%\label{table_water_pos}      % is used to refer this table in the text
\centering                          % used for centering table
\begin{tabular}{lcccccc}        % centered columns (4 columns)
\hline\hline                 % inserts double horizontal lines
IRS	& Group & offset X$^a$  & offset Y$^a$  &	Peak $S_{\nu}$	 & Peak $V_{LSR}$	& Velocity Range \\   
                      &       &       [$\arcsec$] &     [$\arcsec$]  & [Jy] &  [\kms]  &   [\kms]  \\
\hline                        
11 & 26 & -4.594 &-79.578 &10.02 &-57.07 &   -56.22 to -57.18 \\
11 & 27 & -2.797 &-80.876 &19.98 &-57.07 & -56.33 to -57.49 \\
11 & 28 & -2.871 &-81.223 &2.77 &-56.44 &   -56.44 to -56.97 \\
11 & 29 & -2.882 &-81.287 &3.08 &-56.97 &   -56.97 to -57.39 \\
11 & 30 & -2.879 &-81.252 &20.05 &-56.97 & -56.44 to -57.49  \\
11 & 31 & -2.839 &-81.272 &4.61 &-56.86 &   -56.54 to -56.97   \\
11 & 32 & -2.787 &-80.850 &7.56 &-57.18 &  -56.65 to -57.39 \\
11 & 33 & -2.761 &-80.925 &3.04 &-56.97 &    -56.65 to -56.97 \\
11 & 34 & -2.970 &-80.858 &11.43 &-57.81 & -56.97 to -57.81 \\
11 & 35 & -2.763 &-80.905 &3.31 &-57.07 & -57.07 to -57.28 \\
11 & 36 & -2.789 &-80.938 &2.40  & -57.28 &-57.07 to -57.28 \\
11 & 37 & -2.840 &-81.266 &2.72 &-57.07 & -57.07 to -57.28  \\
11 & 38 & -2.952 &-80.824 &3.21 &-57.28 &-57.28 to -57.70 \\
11 & 39 & -4.240 &-79.974 &81.11 &-58.02 & -57.49 to -58.76 \\
11 & 40 & -4.212 &-80.030 &38.97 &-57.91 & -57.60 to -58.23 \\
11 & 41 & -4.222 &-79.902 &22.69  & -58.02 & -57.60 to -58.55 \\
11 & 42 & -4.227 &-79.945 &36.30  & -58.02 & -57.70 to -58.55 \\
11 & 43 & -4.173 &-79.965 &22.74  & -58.23 & -57.81 to -58.44 \\
11 & 44 & -4.233 &-80.036 &13.97 &-58.23 &   -58.12 to -58.65 \\
11 & 45 & -4.242 &-80.017 &34.93 &-58.34 & -57.91 to -58.76 \\
11 & 46 & -4.303 &-79.883 &15.34 &-58.02 &   -57.91 to -58.34 \\
11 & 47 & -3.068 &-80.774 &28.26 &-60.23 &  -59.92 to -60.76 \\
11 & 48 & -3.055 &-80.732 &15.82 &-60.34 & -60.13 to -60.44 \\
11 & 49 & -3.038 &-80.696 &9.06  & -60.34 & -60.34 to -60.55 \\
11 & 50 & -3.123 &-80.708 &5.96 &-61.81 & -61.49 to -62.02 \\
11 & 51 & -3.115 &-80.665 &2.34 &-61.71  &-61.60 to -61.92 \\
11 & 52 & -3.172 &-80.408 &1.77 &-62.23 &   -62.23 to -62.55 \\
11 & 53 & -3.162 &-80.369 &2.63  & -62.23 &  -62.23 to -62.55 \\
11 & 54 & -2.800 &-80.898 &50.61 &-63.81 & -63.71 to -64.34 \\
11 & 55 & -2.802 &-80.879 &245.82 &-63.39 &-62.44 to -64.44 \\
11 & 56 & -2.791 &-80.843 &221.66 &-63.18 &-62.44 to -65.08 \\
11 & 57 & -2.767 &-80.918 &114.46 &-63.18 &  -62.65 to -63.19 \\
11 & 58 & -2.826 &-80.779 &22.41  & -63.18 &  -62.76 to -63.29 \\
11 & 59 & -2.783 &-80.800 &54.47 &-63.39 & -62.86 to -63.71 \\
11 & 60 & -2.791 &-80.934 &159.10  & -63.30 & -62.76 to -64.13 \\
11 & 61 & -2.877 &-80.791 &107.47  & -63.60 & -62.86 to -64.02 \\
11 & 62 & -2.734 &-80.850 &3.92 &-63.30 & -63.39 to -63.60 \\
11 & 63 & -2.734 &-80.857 &128.98 &-63.39 &   -62.97 to -64.02 \\
11 & 64 & -3.175 &-80.372 &50.63 &-63.50 &   -63.39 to -64.87 \\
11 & 65 & -3.177 &-80.414 &25.18  & -64.04 & -63.60 to -64.55 \\
11 & 66 & -2.775 &-80.799 &3.42 &-63.92 & -63.92 to -64.13 \\
11 & 67 & -3.165 &-80.336 &5.61 &-64.34  &-64.13 to -64.76 \\
11 & 68 & -2.778 &-80.815 &2.26 &-64.34 &   -64.23 to -64.66 \\
New & 1 & -11.957 & -21.192 &19.79 &-46.12 & -45.17 to -47.06 \\
New & 2 & -11.960 &-21.199 &3.81 &-52.96 & -52.75 to -53.28 \\
New & 3 & -11.980 &-21.215 &8.31 &-56.12  &-55.49 to -56.86 \\
New & 4 & -13.992 &-23.778 &28.29 &-57.81 & -57.07 to -58.02 \\
New & 5 & -13.989 &-23.756 &198.18 &-57.60 &  -56.33 to -59.49 \\
New & 6 & -13.964 &-23.785 &2.68 &-56.97 &-56.65 to -56.97 \\
New & 7 & -11.997 &-21.222 &35.84  & -63.39 & -61.39 to -63.39 \\
\hline                                  
\end{tabular}
\tablefoot{$^a$ All offsets are relative to the position RA=23h13m45.364s, DEC= +61$^{\circ}$28'10.550"). The uncertainty is less than 0.003$\arcsec$.}
%\vspace{30 mm}
%\end{Large}
\end{table*}

\end{appendix}

\end{document}